\documentstyle[pra,aps]{revtex}

\def\beq{\begin{equation}}
\def\eeq{\end{equation}}

\def\reff#1{(\ref{#1})}

\def\halb{\frac{1}{2}}

\def\crit{{\mbox{\rm\scriptsize crit}}}

\def\energy{{\cal{E}}}
\def\energyres{{\cal{E}}_{\mbox{\rm\scriptsize res}}}

\def\imagi{\mbox{\rm i}}
\def\diff{\,\mbox{\rm d}}

\begin{document}
\draft

\title{Ejection Energy of Photoelectrons in Strong Field Ionization}
\author{D.~Bauer}
\address{Theoretical Quantum Electronics (TQE)\cite{www}, Technische Hochschule
Darmstadt,\\ Hochschulstr.\ 4A, D-64289 Darmstadt, Germany\\
\parbox{8cm}{\begin{center}
{\sc To be published in Phys.\ Rev.\ A.} \\ \mbox{\copyright} The
American Physical Society 1996. \\ All rights reserved.\\ Except as provided under U.S.\ copyright law, this work may not be
reproduced, resold, distributed or modified without the express permission of
The American Physical Society.  The archival version of this work was published
in .................
 \end{center}}}

\date{\today}

\maketitle

\begin{abstract}
We show that zero ejection energy of the photoelectrons is classically
impossible for hydrogen-like ions, even when field 
ionization occurs adiabatically. To prove this we transform the basic equations to those
describing two 2D anharmonic oscillators. The same method yields an
alternative way to derive the anomalous critical field of
hydrogen-like ions. The analytical results are confirmed and
illustrated by numerical simulations. 
\end{abstract}

\pacs{PACS Number(s): 32.80.Rm}

\section{Introduction}
A wealth of new phenomena in the interaction of strong, short laser pulses with
matter has been found in recent years
\cite{piraux,more}. Laser systems delivering pulses at irradiances up to
$10^{19}$~Wcm$^{-2}\mu$m$^2$ are available in several laboratories.
Pulses not longer than a few tens of femtoseconds are ``state of the
art''. In describing their interaction with matter usual perturbation theory fails since the electric field of
such laser pulses is of the order of the atomic field. 
One of the prominent phenomena occuring in a strong laser pulse is
field ionization.
Here the electron is able to escape over the barrier formed by the
Coulomb potential and the laser field. This process is much faster
than tunneling which is dominant in weaker fields. 
In other words, in intense laser fields  the so-called barrier suppression ionization (BSI)
regime \cite{augst} is reached. In this regime, classical and semi-classical
pictures are expected to work well and much theoretical research has been done in that
direction, both analytically \cite{krainov,fedorov,grochmal} and
numerically \cite{kyrala,keitel,lerner_i,lerner_ii,lerner_iii}.  

So-called ``simple man's theory'' (SMT) \cite{vanLinden,gallagh,corkum} succeeded in
explaining essential features of above threshold ionization
(ATI) (see e.g.\ \cite{piraux,freeman} for an overview), 
such as cut-offs in photoelectron energies and harmonic spectra
\cite{Krause}. 
Extensions to SMT considering rescattering effects
clarified experimentally observed features such as plateaus in
ATI-spectra \cite{paulus} and rings in the angular distributions \cite{yang}.

However, no
ionization dynamics enters in the SMT since the electron is regarded
as being ``born'' in the laser field without perceiving any attraction
by the ionic
core anymore, at least until rescattering by the nucleus. Even in
Keldysh-Faisal-Reiss (KFR)-type theories \cite{keldysh,faisal,reiss}  the Coulomb potential
enters not directly but through the initial energy state wave function.

For relevant applications, e.g.\ inverse bremsstrahlung, it is
important to know the energy with which the electron is freed
(ejection energy) \cite{mulser}. To calculate the latter the
ionization dynamics must be described by solving the time-dependent
Schr\"odinger equation or by using simpler alternative concepts. In
this paper a classical model is used for this purpose.

For the sake of completeness, we briefly review SMT.
Suppose that a linearly polarized laser field $E(t)$ sets in at $t=0$, the
velocity of a non-relativistic  electron released at time $t'$ is at
the instant $t$
\beq v(t)=-\int_{t'}^t E(\tau) \diff \tau + v_{00} = -\int_0^t  E(\tau)
\diff \tau + v_0(t') + v_{00} \label{smtbasics} \eeq 
(atomic units will be used throughout this paper) with
\[ v_0(t')=-\int_{t'}^{0} E(\tau) \diff \tau \]
the residual drift due to ejection out of phase, i.e.\ {\em not} in
the maxima of the field. $v_{00}$ is the ejection velocity we are
particularly interested in. Since the ejection velocity $v_{00}$
contributes to the overall residual kinetic energy $\energyres$ of the electron at
the end of the pulse,
\[ \energyres=\halb m (v_0 + v_{00})^2, \]
it alters the electron distribution function and must be taken into account.
Ignoring $v_{00}$ in \reff{smtbasics} leads in the case of linear
polarized laser light to the well-known results
$\langle \energyres \rangle= 3 U_p$, $\max{\energyres}=8U_p$ as well as to
the maximum return energy $\approx 3.2 U_p$ which is responsible for
the cutoff in harmonic-spectra. $U_p$ is the ponderomotive potential
$e^2 E^2/(4m\omega^2)$, i.e.\ the mean quiver energy of the electron
in the field.

The paper is organized as follows: in Sec.~\ref{trafo} the anharmonic
oscillator picture is introduced which leads to a physically appealing
interpretation of the adiabatic ionization process as discussed in
Sec.~\ref{picture}. In Sec.~\ref{limits}, the lower limit for the ejection energy as well
as the upper limit for the ejection radius is derived. It is further
shown that no upper limit for the ejection energy exists. In
Sec.~\ref{numerics} the analytical results are confirmed and illustrated by
numerical simulations. Sec.~\ref{quantum} is devoted to a rederivation of the
anomalous critical field of hydrogen like ions \cite{shakeshaft} in
the framework of the anharmonic oscillator picture. The adiabatic
deformation of the initial ground state is calculated and compared
with the numerical solution of the time-dependent
Schr\"odinger equation.
Finally, we give our conclusion in Sec.~\ref{concl}.

\section{Transforming to anharmonic oscillators} \label{trafo}
The laser frequency $\Omega$ can be regarded as small compared to the
orbital frequency $\omega$ of the bound electron (e.g.\ $\Omega=0.18$~a.u.\ for KrF
and $\Omega=0.04$~a.u.\ for Nd). Therefore, a quasi adiabatic
treatment is appropriate. We will now outline the transformation
to a set of two 2D anharmonic oscillators \cite{Fujikawa}.

We start with the Hamiltonian describing an electron in a Coulomb
potential $-Z/\sqrt{\rho^2+z^2}$, and a static electric field $E$ aligned
in $z$ direction,

\beq H(\rho,z;p_\rho,p_z,p_\varphi)=\halb \left( p_z^2 + p_\rho^2 +
  \frac{p_\varphi^2}{\rho^2}\right) -\frac{Z}{\sqrt{\rho^2+z^2}} + Ez
= \energy. \label{cylhamilt} \eeq
Here, $\energy$ is the total energy of this conservative system and
the azimuthal momentum $p_\varphi$ is a constant of the motion.

We transform to parabolic coordinates first. 
The new coordinates $\xi$ and $\eta$ are related to the cylindric ones according
to 
\[ \xi=( r-z)/2, \quad \eta=(r + z)/2,\] 
with $r=\sqrt{\rho^2+z^2}$. 
The canonical momenta are related through
\[ p_\xi=\sqrt{\frac{\eta}{\xi}}p_\rho-p_z, \quad
p_\eta=\sqrt{\frac{\xi}{\eta}}p_\rho+p_z. \] 
This leads to the Hamiltonian in parabolic coordinates,
\beq H(\xi,\eta;p_\xi,p_\eta,p_\varphi)=\halb\left(
  \frac{\xi}{\xi+\eta}p_\xi^2 + \frac{\eta}{\xi+\eta}p_\eta^2 +
  \frac{1}{4\xi\eta} p_\varphi^2\right) -\frac{Z}{\xi+\eta}+
E(\eta-\xi) = \energy. \label{parabolhamilt} \eeq 
It is well known that the
Hamilton-Jacobi-Equation of the problem
separates in parabolic coordinates \cite{landau}, as does so
Schr\"odinger's equation. It is advantageous to perform another
canonical transformation \cite{Fujikawa}, 
\[ \xi = u^2/4,\quad \eta=v^2/4, \quad 0\leq
u,v < \infty, \]
\[ p_u=\sqrt{\xi}p_\xi,\quad  p_v=\sqrt{\eta} p_\eta,\]
leading to
\beq
H(u,v;p_u,p_v,p_\varphi)=\halb\left(\frac{4}{u^2+v^2}\right)\left(
  p_u^2 + \frac{1}{u^2}p_\varphi^2 + p_v^2 +
  \frac{1}{v^2}p_\varphi^2\right)
-\frac{4Z}{u^2+v^2}+g(v^2-u^2)=\energy, \label{uvhamilt} \eeq
where the new field $g = E/4 $
has been introduced.

Now, we define the ``zero energy Hamiltonian'' $ H_0=H-\energy\equiv 0$, 
and by multiplying this equation with $(u^2+v^2)/4=r$ we are left with
\[ H'_0=\halb\left( p_u^2 + \frac{p_\varphi^2}{u^2} + p_v^2 +
  \frac{p_\varphi^2}{v^2}\right) + \frac{\omega^2}{2} (u^2+v^2) - Z +
\frac{1}{4}g(v^4-u^4) \equiv 0, \]
where $\omega^2=-\energy/2$ .
Therefore, we finally get 
\beq H_u(u;p_u,p_\varphi)= \halb\left( p_u^2 +
  \frac{p_\varphi^2}{u^2}\right)+ \frac{\omega^2}{2} u^2-
\frac{1}{4}gu^4 = A, \label{ionizing} \eeq
\beq H_v(v;p_v,p_{\varphi'})= \halb\left( p_v^2 +
  \frac{p_{\varphi'}^2}{v^2}\right)+ \frac{\omega^2}{2} v^2+
\frac{1}{4}gv^4 = B, \label{trapping} \eeq
\beq A+B=Z,\label{aplusb} \eeq
\beq p_\varphi=p_{\varphi'},\label{momenta} \eeq 
which represents a set of two 2D anharmonic oscillators moving
independently. There are only constraints concerning the {\em initial}
values of the system: (i) the energies must sum up to the given
total ``energy'' $Z$, and (ii) the angular momenta, which are constants of
the motion, are equal.

\section{Physical picture} \label{picture}
The physical interpretation of Eqs.\reff{ionizing} and \reff{trapping}
is very simple. Supposing $g\geq 0$, Eq.\reff{trapping} allows only for a
bound motion in $v$, whereas the effective potential in
Eq.\reff{ionizing},
$ V_u(u)=p_\varphi^2/(2 u^2) + \omega^2 u^2/2-gu^4/4, $
 has a local maximum so that the motion in $u$ is
unbound if the energy $A$ lies above the potential barrier.  

Now, the field $g$ may be turned on
adiabatically. Initially, when $g=0$ holds, the $v$- and the
$u$-motion take place in equal potentials. While $g$ increases, the
$u$-potential is bent down whereas the $v$-potential is
steepened. Therefore one expects an adiabatic lowering of the energy $A$ and an
adiabatic raising of the level $B$ (Fig.~\ref{oscillators}). But the frequency $\omega$ may
also change because the adiabatic invariant
\beq S_u(\omega,g,A)=\sqrt{2} \oint \sqrt{A-V_u(u)}\diff u = \mbox{const.} \label{adiainv} \eeq
yields a relation between $g$, $A$ {\em and} $\omega$ only. Indeed, a change
in $\omega$ is the classical dc Stark-shift, of course.
The trick is to consider the special Kepler orbit which ionizes
earliest in an adiabatically ramped dc field. Therefore, we switch
back to the ``physical space'' for a moment:
the classical trajectory which ionizes first is that one mostly
directed towards the potential barrier. There are two Kepler orbits
lying on the $z$-axis in the {\em limit} eccentricity $\to 1$. One lies in
the region $z\geq 0$, the other one in $z\leq 0$. If the potential
barrier comes from negative $z$, as in our case where $g\geq 0$, the
Kepler orbit $z\leq 0$ ionizes first and the Kepler orbit $z\geq 0$
does last. 
The two $z$-directed orbits are the extreme Stark-shifted ones. The
orbit which ionizes first corresponds to the electron with smallest
ejection energy. Therefore, the ejection energy of just this
electron provides a lower limit for all ejection energies which may
occur. Fortunately, this special energy can be calculated.

Now we turn back to the anharmonic oscillator picture:
motion along the negative $z$-axis means $v=p_\varphi=p_{\varphi'}=0$ for all
times. Therefore,
the anharmonic oscillator $H_v$ is frozen, i.e.,
$B\equiv 0$. Now, when $g$ is adiabatically ramped $B$ is {\em not}
adiabatically raised
since the $v$-particle {\em rests} in the potential well and 
does not recognize the steepening of the potential. It 
follows that $B$ remains zero, and $A\equiv Z$ (Fig.~\ref{oscillators_ii}). The
adiabatic invariant \reff{adiainv} now provides a relation between
$\omega$ and $g$ only, i.e., one can, in principal, calculate the
classical Stark-shift $\omega^2(g)$ for this particular orbit. The
Stark-shift is the key which enables us to calculate the lower limit
for the ejection energy and the upper limit for the ejection radius,
as we will see in the following Section. 

\section{Calculation of the ejection energy} \label{limits}
Since the integral in Eq.\reff{adiainv} can not be solved
analytically some approximations must be applied. We will restrict
ourselves to the linear Stark-shift which follows from
Eq.\reff{adiainv} when the integrand and the turning point is expanded
in powers of the field $g$. The result corresponds to the quantum
mechanical one in the limit of high quantum numbers, $n\to\infty$. 

The potential barrier for $H_u$ in the case $p_\varphi\equiv 0$ is
located at
$u_b=\omega/\sqrt{g}$.  
The value of the potential at that point is
$ V_u(u_b)=\omega^4/(4g)$.
Since the kinetic energy is zero when the potential barrier meets the
total energy $A=Z$ for the first time (this is the case due to
adiabaticity), one has
\beq \frac{\omega^4}{4g}=A=Z \label{relation_one} \eeq
at that certain moment. 

Now, we assume a linear dependence between energy and field, i.e.,
only the linear Stark effect is included in our further calculations,
\beq \energy(E)=\energy_0+\frac{4}{\lambda} E, \qquad \lambda<0. \label{linstark} \eeq
As derived by expanding the invariant \reff{adiainv}  or from the
classical limit of the well-known quantum mechanical result \cite{landau} 
\[ \lambda=\frac{16\energy_0}{3 Z} \]  
holds, which will be used in the following. 

The ``over the top''-criterion \reff{relation_one} then reads
\[ \frac{\energy^2}{\lambda (\energy-\energy_0)}=Z, \]
and this leads to the
``over the top''-field strength 
\beq E_b=E(\energy_b)=\frac{4}{9}\frac{\energy_0^2}{Z}. \label{overtopfield} \eeq 
The electron is not yet free when it flows over the barrier because its Coulomb energy
$\energy-Ez$ is still negative. Demanding $\energy-Ez=0$,
which may be written in $u$ and $v$ as $\energy-g(v^2-u^2) = 0$,
is equivalent to claim
\[ V_u(u)=\frac{\omega^2}{2}u^2-\frac{1}{4}gu^4 =0\]
when $v=0$, as it is in our case. Therefore
$ p_u^2/2 = A = Z $
holds during the moment when the electron becomes free.
The coordinate $u_0$ where this happens is easily determined through
$u_0^2/4=\omega^2/(2g)$
This equation gives an {\em upper} limit for the ejection radius,
\beq r_0=3 \frac{Z}{\vert\energy_0\vert}. \label{upperlimitradius}\eeq
The physical kinetic energy $T$ is related to $p_u$ according
Eq.\reff{uvhamilt} (using $p_v=v=p_\varphi=0$) through
$ T:=2 p_u^2/u^2 $.
For $u_0$ this leads to
$ T_0=2Zg/\omega^2$.
Going back to the physical quantities $\energy$ and $E$ instead of
$\omega^2$ and $g$ this means 
$ T_0=-ZE/\energy$.
Assuming that $\energy$ and $E$ do not vary much during the short time
between ``flowing over the top'' and ``getting positive Coulomb energy'',
i.e.\ ionization, and inserting $\lambda$, we get
\beq T_0=\frac{1}{3}\vert\energy_0\vert. \label{ejectenergfinal} \eeq

Note that the ejection energy is essentially
the kinetic energy gained by the ``$u$-particle'' when it slides from
the top of the potential barrier down to $V_u(u_0)=0$. Multiplication
with a factor $4/u_0^2$ leads to the physical ejection energy $T_0$.

We have already mentioned the two extreme type
of orbits: they are both aligned along the $z$-axis (eccentricity $\to 1$), the early
ionizing one on the negative half, the latest ionizing orbit on the
positive half. For the latter one $u=0$ and $A=0$ holds instead of
$v=0$ and $B=0$. Therefore, we have only to
substitute $g\to -g$ in the result for the Stark-shift. 
Obviously, there is no potential barrier in $V_v(v)$ over which the
electron could escape. Nevertheless, from the ``zero Coulomb
energy''-condition $V_v(v)=0$ follows, as in the $u$-case. This
condition can only be fulfilled through $v=0$. Thus, $u=v=0$ holds
simultaneously, i.e., the electron ionizes over the Coulomb
singularity in the limit eccentricity $\to 1$, and the kinetic energy
is infinite at that point. So we conclude that there is no {\em upper}
limit for the ejection energy.

\section{Numerical results} \label{numerics}
Our numerical simulations presented in the following were performed for the ``classical'' 1s state of
atomic hydrogen, $Z=1$, $\energy_0=-1/2$.
The classical linear Stark effect then gives
$\lambda = -8/3$.
The ``over the
top''-values are calculated to be
$\energy_b=-2/3$, $E_b=1/9$,  
and, finally, the lower limit for the ejection energy is found to be
$T_0=1/6$.
Eq.\reff{upperlimitradius} gives for the ejection radius
$r_0=\vert z_0\vert=6$, i.e., all electrons will be ejected within a
sphere of radius 6~a.u..

Numerically, the exact motion of an ensemble of electrons in a Coulomb potential
and an electric field can easily be determined using  
the so-called ``classical trajectory Monte Carlo-method''
(CTMC) \cite{cohen}.
We have 
performed CTMC runs and looked at each test electron when its
Coulomb energy $\energy-E(t)z$ becomes positive, i.e., when the
electron would be free if the field is turned off immediately (this is
our definition of ionization). 

In Fig.\ref{fig_ejectionenergy} the
kinetic ejection energy of each electron is indicated by a symbol over
the field strength where ionization takes place (the field is
directly proportional to the time due to linear ramping) for 
$ Z=1$, $\energy_0=-1/2$, $ E(t)=t/2000$,
and an ensemble consisting of 1000 test electrons. The lower limit of
the ejection energy is indicated in the plot by a horizontal line. 

In Fig.\ref{fig_ejectionradius} a similar plot for the ejection radius
is presented.

An oscillating field $E(t)\sim\cos\Omega t$ can be treated adiabatically
when the frequency $\Omega$ is small compared to the orbital
frequency $\vert\energy_0\vert/\hbar$, i.e.\ in atomic units
$ \Omega/\vert\energy_0\vert \ll 1$.
We repeated our numerical run discussed in the previous paragraph with
the field replaced by
$ E(t)=t\cos(\Omega t)/2000$, $\Omega=0.04$.
The frequency $\Omega$ corresponds to Nd laser-light and due to
$\Omega/\vert\energy_0\vert=0.08$ an almost adiabatic behavior is
expected.
In Fig.\ref{oscill_energies} the kinetic ejection energies of each
test particle is plotted vs.\ its ionization time. The absolute value
of the electric field and the calculated lower limit for the ejection
energy are included. Apart from
three test particles which are released extremely out of phase, i.e.,
when the electric field is low, all electrons lie above the calculated
lower limit.

\section{Quantum mechanical calculations} \label{quantum}
The anharmonic oscillator method provides also an elegant and alternative way
to derive the anomalous critical field \cite{shakeshaft} for hydrogen-like ions. With
the ansatz $ \psi(u,v,\varphi,\varphi'(\varphi))=\psi_u(u)\exp(\imagi m
\varphi)\psi_v(v)\exp(\imagi m' \varphi')$ the Hamiltonians
\reff{ionizing} and \reff{trapping} lead to the following two
Schr\"odinger equations
\[ \left\{ -\halb \left[ \frac{1}{u} \partial_u ( u \partial_u) -
    \frac{m}{u^2}\right] + \frac{\omega^2}{2} u^2 - \frac{1}{4} g u^4
\right\} \psi_u(u) = A \psi_u(u), \]
\[ \left\{ -\halb \left[ \frac{1}{v} \partial_v ( v \partial_v) -
    \frac{m'}{v^2}\right] + \frac{\omega^2}{2} v^2 + \frac{1}{4} g v^4
\right\} \psi_v(v) = B \psi_v(v), \]
\[ m=m', \] 
\[ A+B=Z. \] 
Since the solution of the unperturbed problem, which is a 2D harmonic
oscillator,  is a Gaussian in the $m=m'=0$-case, we use 
\[ \psi_u(u)=\sqrt{a_u/\pi} \exp(-a_u u^2/2) \] 
as a trial function.
The total energy then is
\[ A(g)=2 \pi \int_0^\infty \diff u\, u \psi^*_u H_u \psi_u =
\frac{1}{a_u} \left( \frac{\omega^2}{2} - \frac{a_u^2}{2} \right) +
a_u - \frac{g}{2 a_u^2}. \]
Minimizing this energy yields up to first order in $g$ $a_u=\omega
(1-g/\omega^3)$ and $a_v=\omega (1+g/\omega^3)$. The oscillator
energies are $A(g)=\omega(1-g/(2\omega^3))$ and
$B(g)=\omega(1+g/(2\omega^3))$. Note that this is consistent with the
fact that the linear Stark-effect vanishes for the ground state of
hydrogen-like ions, since $A(g)+B(g)=2\omega$ and from $A+B=Z$ and
$\omega^2=-\energy_0/2$ follows $\energy_0=-Z^2/2$. 
For the hydrogen 1s-state exposed to an adiabatically ramped dc field
$E$, the wave function
in physical coordinates is
\beq \psi_H(\rho,z)= \frac{1-4E^2}{\sqrt{\pi}}
\exp(-\sqrt{\rho^2+z^2})\exp(- 2 E z), \label{deform} \eeq
i.e., the unperturbed wave function is multiplied by a ``deformation
factor'' $\exp(-2Ez)$. 

In order to calculate the critical field $E_\crit=4 g_\crit$ we claim
that the $u$-Gaussian touches the potential barrier energetically:
\[ A(g)=\omega - \frac{g}{2\omega^2} = \frac{\omega^4}{4 g}. \]
This leads to
\[ g_\crit=\omega^3 ( 1-1/\sqrt{2} ) \approx 0.3 \omega^3 . \]
For the 1s hydrogen state we get
$ E^H_\crit=0.15\mbox{ a.u.} $
in accordance with the result presented in \cite{shakeshaft}.

In Fig.~\ref{contcomp} a comparison is made between the calculated deformation
$\vert \psi_H(\rho,z)\vert^2$ which results from Eq.~\ref{deform} and
the exact numerical solution of the time-dependent
Schr\"odinger equation where we ramped the field linearly over 30
atomic time units up to the critical field  $ E^H_\crit=0.15\mbox{
a.u.} $. Strong deviation occurs only for the 1\%-contour while the
main parts of the probability densities agree well. 

In Fig.~\ref{starkpotpic} the initial probability density and the variationally
determined one at the critical field are shown in the energy diagram. 
Note that at the critical
field the packet lies $0.27$ atomic energy units above
the barrier in $z$-direction. This value may be interpreted as a
quantum mechanical ejection energy.

\section{Conclusions} \label{concl}
We have  shown that within the framework of the anharmonic
oscillator model important features in strong field
ionization can be derived. Using this method, we have  calculated (i) the
lower limit for the ejection energy, $T_0=\vert \energy_0 \vert /3$, (ii) the upper limit for the
ejection radius, $ r_0=3Z/\vert\energy_0\vert $, and (iii) the anomalous critical field for
hydrogen-like ions as well as the shape of the deformed probability
density. We have  also demonstrated that there exists no upper limit
for the ejection energy. All results have been confirmed by CTMC simulations.

\begin{figure}

\caption{\label{oscillators}} The two 2D anharmonic oscillators in the
case $p_\varphi=p_{\varphi'}=0$. Initially, when $g=0$ holds the two
potentials are pure harmonic ones (dashed). As $g$ is adiabatically
raised one expects an adiabatic lowering of level $A$ and a raising
of level $B$. 
\end{figure}

\begin{figure}
\caption{\label{oscillators_ii}} For the early ionizing Kepler orbit
discussed in the text $v\equiv 0$ holds. Therefore, and owing to the
constraint $A+B=Z$, both levels do not move when $g$ is adiabatically raised.
\end{figure}

\begin{figure}
\caption{\label{fig_ejectionenergy}} The ejection energy of all test electrons in $z$- and
  $\rho$ direction ($\Diamond,\bigtriangleup$), and the sum of both (\mbox{\rule{2mm}{2mm}}).
The calculated lower limit for the ejection energy is also
  shown. One can see that this limit is confirmed by the numerical runs: none
  of the kinetic energies \mbox{\rule{2mm}{2mm}} lies beneath the limit. 
\end{figure}

\begin{figure}
\caption{\label{fig_ejectionradius}} The radii of all test electron
  when they become free. The analytically calculated upper limit is
  confirmed by the numerical run. All electrons are ionized within
  a sphere of radius 6 atomic units.    
\end{figure}

\begin{figure}
\caption{\label{oscill_energies}} The kinetic ejection energies in an
  oscillatory $E$-field. Only 3 of 1000 test electrons ionize through
  a ``non-adiabatic channel'' in such a way that they lie
  beneath the calculated lower limit. Note, that these particles
  become free when the field is low. The absolute value of the
  electric field is also shown.
\end{figure}

\begin{figure}
\caption{\label{contcomp}} The contour lines representing 80, 50, 10 and 1\% of
  the peak height at the origin $\rho=z=0$ are shown for the numerical
  result (solid) and the variationally calculated probability density (dotted). The
  field strength is the critical one $E=0.15$~a.u..
\end{figure}

\begin{figure}
\caption{\label{starkpotpic}} Unperturbed and Stark-deformed 1s peak
  for atomic hydrogen at the critical field strength
  $E_\crit=0.15$~a.u.. The energy gap of 0.27~a.u.\ between the energy
  level at $\energy_0=-1/2$ and the barrier can be interpreted as an
  quantum mechanical ejection energy.
\end{figure}

\end{document}